\documentclass[onecolumn,showpacs]{revtex4}

\usepackage{graphicx}
\usepackage{dcolumn}
\usepackage{bm}

\input epsf

\begin{document}

\title{\Large Role of Pressure in Quasi-Spherical Gravitational Collapse}

\author{\bf Subenoy Chakraborty}
\email{subenoyc@yahoo.co.in}
\author{\bf Sanjukta Chakraborty}
\author{\bf Ujjal Debnath}
\email{ujjaldebnath@yahoo.com}

\affiliation{Department of Mathematics, Jadavpur University,
Calcutta-32, India.}

\date{\today}

\begin{abstract}
We study quasi-spherical Szekeres space-time (which possess no
killing vectors) for perfect fluid, matter with tangential stress
only and matter with anisotropic pressure respectively. In the
first two cases cosmological solutions have been obtained and
their asymptotic behaviour have been examined while for
anisotropic pressure, gravitational collapse has been studied and
the role of the pressure has been discussed.
\end{abstract}

\pacs{04.20, 04.20 Dw, 04.70 B}

\maketitle

\section{\normalsize\bf{Introduction}}
In cosmology, solutions to Einstein's field equations are obtained
by imposing symmetries [1] on space-time. Usually, spatial
homogeneity is one of the reasonable assumptions (in an average
sense) while for cosmological phenomena over galactic scale or in
smaller scale, inhomogeneous solutions are useful. Szekeres [2] in
1975 gave a class of inhomogeneous solutions representing
irrotational dust. The space-time represented by these solutions
has no killing vectors and it has invariant family of spherical
hypersurfaces. Hence this space-time is referred as quasi-spherical space-time.\\

 An extensive study of gravitational collapse [3-8] has been carried
 out of Tolman-Bondi-Lema\^{\i}tre (TBL) spherically symmetric
 space-times containing irrotational dust to support or disprove
 the cosmic censorship conjecture (CCC). A general conclusion from
 these studies is that a central curvature singularity forms but
 its local or global visibility depends on the initial data. Also
 over the past decades, the role of pressure within a collapsing
 cloud has been studied [9-13] but the actual role of pressure in
 determining the end state of a continual collapse is not yet
 clear.\\

 On the other hand, there is very little progress in studying
 non-spherical collapse. Basically, the difficulty is the
 ambiguity of horizon formation in non-spherical geometries and
 the influence of gravitational radiation. Though there is hoop
 conjecture by Thorne [14] to characterize the formation of
 horizon but only few works [15-19] have been done to confirm or
 refute the conjecture. Recently, an extensive study of
 irrotational dust collapse has been done in quasi-spherical
 Szekeres space-time both for four [20] and higher [21] dimensions.
 In this paper, we have done an extensive analysis of Szekeres
 model with matter containing pressure and studied the collapsing
 procedure to examine the role of pressure in characterizing the
 final singularity. The paper is organized as follows: The
 Szekeres model [22] has been described in section II. The perfect
 fluid solution with asymptotic behaviour has been presented in
 section III. Section IV deals with the tangential stress only
 while section V deals with gravitational collapse with
 non-isotropic pressure. Finally, the paper ends with a brief
 discussion in section VI.\\

\section{\normalsize\bf{The Szekeres' Model}}
The metric ansatz for the four dimensional Szekeres' space-time
[22] is of the form

\begin{equation}
ds^{2}=dt^{2}-e^{2\alpha}dr^{2}-e^{2\beta}(dx^{2}+dy^{2})
\end{equation}

where $\alpha$ and $\beta$ are functions of all space-time
variables i.e., $$\alpha=\alpha(t,r,x,y),~~
\beta=\beta(t,r,x,y).$$

Considering both radial and transverse stresses the energy
momentum tensor has the following structure
\begin{equation}
T_{\mu}^{\nu}=\text{diag}(\rho,-p_{r},-p_{_{T}},-p_{_{T}})
\end{equation}

Hence the full set of Einstein equations are

\begin{equation}
2\dot{\alpha}\dot{\beta}+\dot{\beta}^{2}-e^{-2\beta}(\alpha_{x}^{2}+
\alpha_{y}^{2}+\alpha_{xx}+\alpha_{yy}+\beta_{xx}+\beta_{yy})+
e^{-2\alpha}(2\alpha'\beta'-3\beta'^{2}-2\beta'')=\rho
\end{equation}

\begin{equation}
 3\dot{\beta}^{2}+2\ddot{\beta}-e^{-2\alpha}\beta'^{2}-e^{-2\beta}
 (\beta_{xx}+\beta_{yy})=-p_{r}
\end{equation}

\begin{equation}
\ddot{\alpha}+\dot{\alpha}^{2}+\ddot{\beta}+\dot{\beta}^{2}+\dot{
\alpha}\dot{\beta}+e^{-2\alpha}(\alpha'\beta'-\beta'^{2}-\beta'')-
e^{-2\beta}(\alpha_{y}^{2}+\alpha_{yy}-\alpha_{y}\beta_{y}+
\alpha_{x}\beta_{x})=-p_{_{T}}
\end{equation}
\begin{equation}
\ddot{\alpha}+\dot{\alpha}^{2}+\ddot{\beta}+\dot{\beta}^{2}+
\dot{\alpha}\dot{\beta}+e^{-2\alpha}(\alpha'\beta'-\beta'^{2}-
\beta'')-e^{-2\beta}(\alpha_{x}^{2}+\alpha_{xx}-\alpha_{x}\beta_{x}
+\alpha_{y}\beta_{y})=-p_{_{T}}
\end{equation}
\begin{equation}
\alpha_{y}(-\alpha_{x}+\beta_{x})+\alpha_{x}\beta_{y}-\alpha_{xy}=0
\end{equation}
\begin{equation}
\dot{\alpha}\beta'-\dot{\beta}\beta'-\dot{\beta'}=0
\end{equation}
\begin{equation}
-\dot{\alpha}\alpha_{x}+\dot{\beta}\alpha_{x}-\dot{\alpha_{x}}-\dot{\beta_{x}}=0
\end{equation}
\begin{equation}
-\dot{\alpha}\alpha_{y}+\dot{\beta}\alpha_{y}-\dot{\alpha_{y}}-\dot{\beta_{y}}=0
\end{equation}
\begin{equation}
\alpha_{x}\beta'-\beta_{x}'=0
\end{equation}
\begin{equation}
\alpha_{y}\beta'-\beta_{y}'=0
\end{equation}
where dot, dash and subscript stands for partial differentiation
with respect to $t$, $r$ and the corresponding variables
respectively (e.g.,
$\beta_{x}=\frac{\partial\beta}{\partial x}$).\\

Combining the time derivatives of equations (11) and (12) with the
differentiation of (8) with respect to $x$ and $y$ separately we
have the integrability condition

\begin{equation}
\beta'\dot{\beta}_{x}=0=\beta'\dot{\beta}_{y}
\end{equation}

Hence we may have three possibilities namely,

\begin{equation}\begin{array}{cc}
 (a)~ \beta'\ne 0,~\dot{\beta_{x}}=0=\dot{\beta_{y}},
\\\\
 (b)~ \beta'=0,~
\dot{\beta_{x}}=0=\dot{\beta_{y}}
\\\\
~~~~ (c)~ \beta'=0,~ \dot{\beta_{x}}\ne 0,~ \dot{\beta_{y}}\ne 0
\end{array}
\end{equation}

Following Szekeres' [22] the field equations are not solvable for
the third case so we shall consider only the first two cases.\\

The energy conservation equation namely $T^{\mu\nu};_{\nu}=0$
gives

\begin{equation}
\dot{\rho}+\rho(\dot{\alpha}+2\dot{\beta})+(\dot{\alpha}p_{r}+2\dot{\beta}p_{_{T}})=0
\end{equation}
\begin{equation}
\frac{\partial}{\partial
x}(p_{_{T}}e^{\alpha})=p_{r}\alpha_{x}e^{\alpha}
\end{equation}
\begin{equation}
\frac{\partial}{\partial
y}(p_{_{T}}e^{\alpha})=p_{r}\alpha_{y}e^{\alpha}
\end{equation}
\begin{equation}
\frac{\partial}{\partial
r}(p_{r}e^{2\beta})=p_{_{T}}\frac{\partial}{\partial
r}(e^{2\beta})
\end{equation}

Now for the first choice namely $\beta'\ne 0,~
\dot{\beta}_{x}=0=\dot{\beta}_{y}$ we have from the field
equations the explicit form of the metric coefficients are as
follows:

\begin{equation}
e^{\beta}=R(t,r)~e^{\nu(r,x,y)}
\end{equation}

\begin{equation}
e^{\alpha}=R'+R~\nu'
\end{equation}

where $R$ and $\nu$ satisfy the following differential equations

\begin{equation}
2R\ddot{R}+\dot{R}^{2}+p_{r}R^{2}=f(r), ~~~~~(f(r)=\text{arbitrary
separation function})
\end{equation}

and
\begin{equation}
e^{-2\nu}(\nu_{xx}+\nu_{yy})=f(r)-1
\end{equation}

Here we have assumed $p_{r}=p_{r}(r,t)$. Equation (21) has the
first integral

\begin{equation}
\dot{R}^{2}=f(r)+\frac{G(r)}{R}-\frac{1}{R}\int p_{r}R^{2}dR
\end{equation}

while one of the possible solutions of equation (22) can be taken
as [24]

\begin{equation}
e^{-\nu}=A(r)(x^{2}+y^{2})+B_{1}(r) x+B_{2}(r) y +D(r)
\end{equation}

where the arbitrary functions $A(r),~B_{1}(r),~B_{2}(r)$ and
$D(r)$ are related as

$$
B_{1}^{2}+B_{2}^{2}-4AD=f(r)-1
$$

and $G(r)$ is an arbitrary function.\\

For the choice ($b$) the metric coefficients are of the form

\begin{equation}
e^{\beta}=R(t)e^{\nu(x,y)}
\end{equation}
\begin{equation}
e^{\alpha}=R(t)\eta(r,x,y)+\mu(t,r)
\end{equation}

Then as before from the field equation (4) we have similar
differential equation in $R$ and $\nu$ as
\begin{equation}
2R\ddot{R}+\dot{R}^{2}+p_{r}R^{2}=K, ~~~~~(K~~ \text{is a
constant})
\end{equation}
\begin{equation}
e^{-2\nu}(\nu_{xx}+\nu_{yy})=K
\end{equation}

with $K$ an arbitrary constant. Also for $\nu$ we choose as in
case (a)

\begin{equation}
e^{-\nu}=P(x^{2}+y^{2})+Q_{1}x+Q_{2}y+S
\end{equation}

where $P,~Q_{1},~Q_{2}$ and $S$ are arbitrary constants restricted
by the relation

$$
Q_{1}^{2}+Q_{2}^{2}-4PS=K
$$

Now to determine the function $\eta$ we have from the field
equation (7)

\begin{equation}
\frac{\partial^{2}(e^{-\nu}\eta)}{\partial x\partial y}=0
\end{equation}

and then from the field equations (5) and (6) we have a possible
solution

\begin{equation}
e^{-\nu}\eta=u(r)(x^{2}+y^{2})+v_{1}(r)x+v_{2}(r)y+w(r)
\end{equation}

where $u(r),~v_{1}(r),~v_{2}(r)$ and $w(r)$ are arbitrary
functions of $r$ alone.\\

Also the differential equation in $\mu$ is

\begin{equation}
R\ddot{\mu}+\dot{R}\dot{\mu}+\mu(\ddot{R}+p_{_{T}}R)+(p_{_{T}}-p_{r})R^{2}\eta=h(r)
\end{equation}

with
$$
h(r)=2(uS+wP)-(v_{1}Q_{1}+v_{2}Q_{2})~.
$$

Now for explicit solutions, we shall consider the following cases
in the next sections:\\

(i)~~the perfect fluid model (i.e., $p_{r}=p_{_{T}}$)\\

(ii)~~the tangential stress only (i.e., $p_{r}=0,~p_{_{T}}\ne
0$)\\

(iii)~~the general case (i.e., $p_{r}\ne 0,~p_{_{T}}\ne 0$).\\

In fact, cosmological solutions are obtained  (in sections III and
IV) for the first two cases respectively while for the third case
collapsing behaviour has been studied and the role of pressure has
been examined.\\

\section{\normalsize\bf{The perfect fluid model}}

In this case due to energy conservation equations (16)-(18) the
isotropic pressure is function of $t$ only i.e., $p=p(t)$
($p_{r}=p_{_{T}}=p$). As there is no restriction on the energy
density so $\rho$ is in general a function of all the 4 variables
i.e., $\rho=\rho(t,r,x,y)$ and hence no equation of state is
imposed.\\

Now for explicit solution according to Szafron [25] and Szafron
and Wainwright [26] we choose
\begin{equation}
p(t)=p_{c}t^{-s}
\end{equation}

($p_{c}$ and $s$ are positive constants) and we have the general
solution for $R$ as [23]

\begin{equation}R^{\frac{3}{2}}=\left\{\begin{array}{lll}

\sqrt{t}\left\{C_{1}J_{\xi}[\frac{2\sqrt{\lambda}}{|s-2|}t^{-\frac{s-2}{2}}]
+C_{2}Y_{\xi}[\frac{2\sqrt{\lambda}}{|s-2|}t^{-\frac{s-2}{2}}]\right\}\\
\\
\sqrt{t}\left\{C_{1}J_{\xi}[\frac{2\sqrt{\lambda}}{|s-2|}t^{-\frac{s-2}{2}}]
+C_{2}J_{-\xi}[\frac{2\sqrt{\lambda}}{|s-2|}t^{-\frac{s-2}{2}}]\right\}\\
\\
C_{1}t^{q_{1}}+C_{2}t^{1-q_{1}}
\end{array}\right.
\end{equation}

according as $\xi$ is an integer, non-integer and $s=2$. Here
$C_{1}$ and $C_{2}$ are arbitrary functions of $r$ and we have
chosen
$$\xi=\frac{1}{s-2},~~\lambda=\frac{3p_{c}}{4},~~q_{1}=\frac{1}{2}(1+\sqrt{1-3p_{c}}).$$

It is to be noted that to derive the above solution we have chosen
$f(r)=0$. Further, if we consider the dust model (i.e., $p=0$)
then the above solution simplifies to $R\propto t^{2/3}$ which is
the form of the scale factor in the usual Friedman model.\\

Now, the physical and kinematical parameters have the following
expressions

\begin{equation}
\rho=\frac{G'(r)+3G\nu'}{R^{2}(R'+R\nu')}-\frac{p_{c}}{t^{s}}
\end{equation}

\begin{equation}
\theta=\frac{R\dot{R}'+3R\dot{R}\nu'+2\dot{R}R'}{R(R'+R\nu')}
\end{equation}

\begin{equation}
\sigma^{2}=\frac{1}{12}\left[\frac{R(Rf'-2R'f)+
(RG'(r)-3R'G)}{\dot{R}R^{2}(R'+R\nu')}\right]^{2}
\end{equation}

The above solution is for the choice ($a$) (see eq.(13)). For the
choice ($b$) (i.e., eq.(14)) the explicit form for $R$ is same as
in equation (34) except here $C_{1}$ and $C_{2}$ are arbitrary
constants and $K=0$. But we note that the differential equation
(30) is not solvable for the above explicit solutions for $R$. The
physical and kinematical parameters have the expressions as

\begin{equation}
\rho=\frac{2(\dot{R}^{2}-K)}{R^{2}}-\frac{2(\ddot{\mu}+\eta
\ddot{R})}{\mu+\eta R}-\frac{p_{c}}{t^{s}}
\end{equation}

\begin{equation}
\theta=\frac{R\dot{\mu}+2\mu\dot{R}+3R\dot{R}\eta}{ R(\mu+\eta R)}
\end{equation}

\begin{equation}
\sigma^{2}=\frac{1}{3}\left[\frac{R\dot{\mu}-\dot{R}\mu}{
R(\mu+\eta R)}\right]^{2}
\end{equation}

$${\normalsize\bf{Asymptotic~ Behaviour}}$$

We shall now discuss the asymptotic behaviour of the above
solutions. The co-ordinates vary over the range :
$t_{0}<t<\infty;~ -\infty<r<\infty;~ -\infty<x,y<\infty$. For both
the choices ($a$) and ($b$) as $p\ge 0,~ p\ne 0$ so we must have
$\frac{1}{2}<q_{1}<1$. So for $s=2$ for large $t$,

\begin{equation}\begin{array}{lll}
R\sim t^{\frac{2q_{1}}{3}} \\\\

\rho\sim t^{-2}
\\\\
p\sim t^{-2}
\\\\
\theta\sim t^{-1}
\\\\
\sigma^{2}\sim t^{-2}
\\\\
\end{array}
\end{equation}

Hence as $t\rightarrow \infty$, ($p,\rho$) fall off faster compare
to ($\theta,\sigma$), while the scale factor $R$ gradually
increases with time. So the model approaches isotropy along fluid
world line as $t\rightarrow \infty$.\\

\section{\normalsize\bf{Model with tangential stresses only}}

 For this model we have from the conservation equation (18)
 $\beta'=0 $ i.e. choice (b) is only possible here. Also from the
 other conservation equations namely equation (16) and (17) we
 have
 \begin{equation}
 p_{_{T}}=A(r,t)e^{-\alpha}
 \end{equation}

 where A(r,t) is an arbitrary function of $r$ and $t$. But for the
 consistency of the differential eq.(30) $ p_{_{T}}$ (as stated
 earlier) must be a function of $r$ and $t$ and hence $\alpha$ should be
 independent of $x$ and $y$. As a consequence, in the solution (26)
 for $\alpha$, $\eta$ must be independent of $x$ and $y$. Thus for the
 solution (32) for $\eta$ we should have
$$
 u(r)=P\eta_{0}(r), ~ v_{1}(r)=Q_{1}\eta_{0}(r),~
 v_{2}(r)=Q_{2}\eta_{0}(r)  ~~ \text{and} ~~   w(r)=S\eta_{0}(r)
$$
 Hence we have $\eta=\eta_{0}(r)$, an arbitrary function of r alone
 and $h(r)=-K\eta_{0}(r)$.\\

 Now, the differential equation for $R$ has the simple form

 \begin{equation}
 \dot{R}^{2}=a_{1}+\frac{a_{2}}{R} ~,~~~~     (a_{1},~ a_{2} ~\text{are~ constants})
 \end{equation}

 which has a parametric solution of the form

 \begin{eqnarray}\left.
 \begin{array}{c}
 R=\frac{a_{2}}{2(-a_{1})}(1-\text{cos}\phi)\\\\
 t_{c}-t=\frac{a_{2}}{2(-a_{1})^{3/2}}(\phi-\text{sin}\phi)\\\\
 \end{array} \right\}~~~~ \text{for} ~~ a_{1}<0~~(0<\phi<2\pi) \nonumber   \\\\
 \left. \begin{array}{c}
 R=\frac{a_{2}}{2a_{1}}(\text{cosh}\phi-1)\\\\
 t_{c}-t=\frac{a_{2}}{2a_{1}^{3/2}}(\text{sinh}\phi-\phi)
 \end{array}\right\}~~~~~~~~~ \text{for} ~~ a_{1}>0~~(\phi>0) \nonumber  \\   \nonumber \\  \nonumber
 \begin{array}{c}
 \text{and}   \\\\
 R=\left(\frac{9a_{2}}{4}\right)^{1/3}(t_{c}-t)^{2/3} ~~~~~~~~~~~~~~~~ \text{for} ~~ a_{1}=0
 \end{array} \nonumber
 \end{eqnarray}
 Here $t_{c}$ is an integration constant that corresponds to the
 time of arrival of each shell to the central singularity.  \\

Choosing the power law solution (i.e. $a_{1}=0$) for $R$ (i.e., if
$R\propto T^{2/3}$ then equation (27) implies that $K=0$) and
assuming $p_{_{T}}=p_{_{0T}}/T^{2},(T=t_{c}-t)$ (i.e., a function
of $T$ alone), it is possible to have a solution for $\mu$ (from
eq.(32)) as

 \begin{equation}\mu(r,t)=\left\{
 \begin{array}{c}
  C_{1}(r)T^{n_{1}}+C_{2}(r)T^{n_{2}},~~~~~~~~~~~~~~~~~~~~~~~~~~~~~~~~~p_{_{0T}}<1/4\\\\
  C_{1}(r)T^{1/6}\text{cos}(k~\text{ln}T)+C_{2}(r)T^{1/6}\text{sin}(k~\text{ln}T),~~~~~p_{_{0T}}>1/4\\\\
 C_{1}(r)T^{1/6}+C_{2}(r)T^{1/6}\text{ln}T,~~~~~~~~~~~~~~~~~~~~~~~~~~~p_{_{0T}}=1/4
 \end{array} \right.
 \end{equation}

 where $C_{1}$ and $C_{2}$ are arbitrary functions of $r$,
 $n_{1}=\frac{1}{6}+\frac{1}{2}\sqrt{1-4p_{_{0T}}}$,
 $n_{2}=\frac{1}{6}-\frac{1}{2}\sqrt{1-4p_{_{0T}}}$,
 $k=\frac{1}{2}\sqrt{4p_{_{0T}}-1}$ and
 $R_{0}=\left(\frac{9a_{2}}{4}\right)^{1/3}$ (Here we have set $\eta=0$, which has no effect on the metric).\\

 Further, the physical and kinematical parameters have the expressions

 \begin{equation}
\rho=\frac{2(\dot{R}^{2}-K)}{R^{2}}-\frac{2(\ddot{\mu}+\eta
\ddot{R})}{\mu+\eta R}-2p_{_{T}}
 \end{equation}
 \begin{equation}
\theta=\frac{R\dot{\mu}+2\mu \dot{R}+3\eta R\dot{R}}{R(\mu+\eta
R)}
 \end{equation}
 \begin{equation}
\sigma^{2}=\frac{1}{3}\left[\frac{\dot{\mu}R-\mu
\dot{R}}{R(\mu+\eta R)}\right]^{2}
 \end{equation}

 It is to be noted that the solution for R does not depend on
 $p_{_{T}}$ so R has same expression for dust model. But for the
 solution of $\mu$ we have only the power law form $T^{2/3}$ (or
 $T^{-1/3}$) when $R$ has Friedmann like behaviour (i.e. $R\sim T^{2/3}$).
 The difference comes in the matter density. For dust model $\rho$
 is a function of all the four co-ordinate variables while in the
 presence of tangential stress $\rho$ is a function of $t$ and $r$
 only. Finally, the asymptotic behaviour for both the model will
 be very similar.\\

 \section{\normalsize\bf{Role of pressure in gravitational collapse}}
 In the general case when both radial and tangential pressures
 are non-zero and distinct then from the Einstein equations they
 can be obtained in compact form as
 \begin{equation}
 \begin{array}{c}
 \rho=\frac{F'}{\zeta^{2}\zeta'}\\\\
 p_{r}=-\frac{\dot{F}}{\zeta^{2}\dot{\zeta}}\\\\
 p_{_{T}}=p_{r}+\frac{\zeta p_{r}'}{2\zeta'}
 \end{array}
 \end{equation}
 where $F(r,t)=Re^{3\nu}(\dot{R}^{2}-f(r))$ and $\zeta=e^{\beta}$.\\

Since $p_{r}$ is regular initially at the centre and blows up at
the singularity so we can choose $p_{r}$ to be of the form:

\begin{equation}
p_{r}=\frac{g(r)}{R^{n}}
\end{equation}

where $g(r)$ is an arbitrary function such that $g(r)\sim r^{n}$
near $r=0$ to make initial matter density non-zero at the centre
$r=0$ and $n$ is any constant. As a consequence, the expressions
for matter density and tangential stress become

\begin{equation}
\rho=\frac{H'+3H\nu'}{R^{2}(R'+R\nu')}
\end{equation}

\begin{equation}
p_{_{T}}=\frac{g(r)}{R^{n}}\left[1-\frac{nR'}{2(R'+R\nu')}\right]+\frac{g'(r)}{2R^{n-1}(R'+R\nu')}~,~~~~(n\ne
3)
\end{equation}

where $H(R,t)=C(r)-\frac{g(r)}{(3-n)}~R^{3-n}$ and $C(r)$ is an
arbitrary integration function. Also the radial velocity of
collapsing shells at a distance $r$ from the centre is given by

\begin{equation}
\dot{R}^{2}=f(r)+\frac{H(R,t)}{R}
\end{equation}

Now if we choose $R=r$ initially then at the beginning of the
collapse the density and the tangential stress have the initial
values

\begin{equation}
\rho_{i}(r,x,y)=\rho_{i}(r,t_{i},x,y)=\frac{c'+3c\nu'}{r^{2}(1+r\nu')}
\end{equation}

\begin{equation}
p_{_{T_{i}}}=p_{_{T}}(t=t_{i})=\frac{g(r)}{r^{n}}\left[1-\frac{n}{2(1+r\nu')}\right]+\frac{g'(r)}{2r^{n-1}(1+r\nu')}
\end{equation}

where $c(r)=H(r,t_{i})=C(r)-\frac{g(r)}{3-n}~r^{3-n}$.\\

Here it is to be noted that for regular initial data $C(r)$ and
$g(r)$ to be $C^{\infty}$ functions and hence we have the
following series expansions

\begin{equation}
\begin{array}{c}
g(r)=\sum_{j=0}^{\infty}g_{j}~r^{n+j}\\\\
C(r)=\sum_{j=0}^{\infty}C_{j}~r^{3+j}\\\\
\rho_{i}(r)=\sum_{j=0}^{\infty}\rho_{j}~r^{j}\\\\
\nu'=\sum_{j=-1}^{\infty}\nu_{j}~r^{j}\\\\
p_{_{T_{i}}}=\sum_{j=0}^{\infty}p_{j}~r^{j}
\end{array}
\end{equation}
where $\nu_{_{-1}}\ge -1$.\\

In these series expansions the coefficients $g_{j}$'s and
$C_{j}$'s are constants while $\rho_{j}$'s, $\nu_{j}$'s and
$p_{j}$'s are functions of $x$ and $y$. These coefficients are
related among themselves through the relations (54) and (55) as
follows:

\begin{equation}\begin{array}{c}
p_{0}=g_{0},~~
p_{1}=g_{1}\left\{1+\frac{1}{2(1+\nu_{_{-1}})}\right\},~~p_{2}=g_{2}
\left\{1+\frac{1}{(1+\nu_{_{-1}})}\right\}-\frac{g_{1}~\nu_{0}}{2(1+\nu_{_{-1}})^{2}},~~....~~....\\\\
\rho_{0}=3c_{0},~~\rho_{1}=\frac{4+3\nu_{_{-1}}}{1+\nu_{_{-1}}}~c_{1},~~\rho_{2}=
\frac{5+3\nu_{_{-1}}}{1+\nu_{_{-1}}}~c_{2}-\frac{\nu_{0}~c_{1}}{(1+\nu_{_{-1}})^{2}},~~....~~....
\end{array}
\end{equation}
$$
\text{or}
$$
\begin{equation}\begin{array}{c}
p_{0}=g_{0}+\frac{g_{1}}{2\nu_{0}},~~
p_{1}=g_{1}\left(1-\frac{\nu_{1}}{2\nu_{0}^{2}}\right)+\frac{g_{2}}{\nu_{0}},~~p_{2}=g_{2}
\left(1-\frac{\nu_{1}}{\nu_{0}^{2}}\right)+\frac{(\nu_{1}^{2}-\nu_{0}\nu_{2})}{2\nu_{0}^{3}}~g_{1}+
\frac{3g_{3}}{2\nu_{0}},~~....~~....\\\\
\rho_{0}=3c_{0}+\frac{c_{1}}{\nu_{0}},~~\rho_{1}=\frac{2c_{2}}{\nu_{0}}+c_{1}\left(3-\frac{\nu_{1}}
{\nu_{0}^{2}}\right),~~\rho_{2}=\frac{3c_{3}}{\nu_{0}}+c_{2}\left(3-\frac{2\nu_{1}}
{\nu_{0}^{2}}\right)+c_{1}\frac{(\nu_{1}^{2}-\nu_{0}\nu_{2})}{\nu_{0}^{3}}
,~~....~~....
\end{array}
\end{equation}

according as $\nu_{_{-1}}>-1$ or $\nu_{_{-1}}=-1$ and
$c_{i}=C_{i}-\frac{g_{i}}{3-n},~~i=0,1,2,...$.\\

The hypersurface $t=t_{s}(r)$ describing the shell focusing
singularity is characterized by

\begin{equation}
R(t_{s}(r),r)=0
\end{equation}

As the differential equation in $R$ (i.e., eq.(53)) is not
solvable so we shall consider only the marginally bound case
(i.e., $f(r)=0$). Hence in this case, the singularity hypersurface
can be written in explicit form as

\begin{equation}
t_{s}(r)-t_{i}=\frac{2r^{3/2}}{3\sqrt{C(r)}}~_{2}F_{1}[\frac{1}{2},b,b+1,\frac{g(r)r^{3-n}}{C(r)(3-n)}]
\end{equation}

where $_{2}F_{1}$ is the usual hypergeometric function with
$b=\frac{3}{2(3-n)}$~.\\

\subsection{\normalsize\bf{Formation of Trapped Surfaces}}

The event horizon of observers at infinity plays an important role
in the nature of the singularity. As formation of event horizon
depends greatly on the computation of null geodesics whose
computation are almost impracticable for the present space-time
geometry, so we consider closely related concept of a trapped
surface (namely a compact space-like 2-surface whose normals on
both sides are future pointing converging null geodesic families).
In fact, if the 2-surface $S_{r,t}$ ($r=$constant, $t=$constant)
is a trapped surface then it and its entire future development lie
behind the event horizon provided the density falls off fast
enough at infinity. So mathematically, if $K^{\mu}$ denotes the
tangent vector field to the null geodesics which is normal to
$S_{r,t}$ then we have
$$
K_{\mu}~K^{\mu}=0,~~K^{\mu}_{~;~\nu}~K^{\nu}=0~.
$$

Now the convergence or divergence of the null geodesics is
characterized by the sign of the invariant $K^{\mu}_{~;~\mu}$
evaluated on the surface $S_{r,t}=0$ (in fact,
$K^{\mu}_{~;~\mu}<0$ indicates convergence while
$K^{\mu}_{~;~\mu}>0$ stands for divergence). It can be shown that
the inward geodesics converges initially and throughout the
collapsing process while the outward geodesics diverges initially
but becomes convergent after a time $t_{ah}(r)$ (time of formation
of apparent horizon) given by
$$
\dot{R}(t_{ah}(r),r)=-\sqrt{1+f(r)}
$$

Now using equations (23) and (50) we have

\begin{equation}
g(r)R^{3-n}(t_{ah}(r),r)-(n-3)R(t_{ah}(r),r)+(n-3)C(r)=0
\end{equation}

or using the explicit solution for $R$ (from eq.(53))

\begin{equation}
t_{ah}(r)-t_{i}=\frac{2r^{3/2}}{3\sqrt{C(r)}}~_{2}F_{1}[\frac{1}{2},b,b+1,\frac{g(r)r^{3-n}}{C(r)(3-n)}]-
\frac{2R^{3/2}(t_{ah},r)}{3\sqrt{C(r)}}~_{2}F_{1}[\frac{1}{2},b,b+1,\frac{g(r)R^{3-n}(t_{ah},r)}{C(r)(3-n)}]
\end{equation}

From the equations (60) and (62) we see that the shell focusing
singularity that appears at $r>0$ is in the future of the apparent
horizon.\\

As we are interested in central shell focusing singularity (at
$r=0$), so its time of occurrence is given by

\begin{equation}\begin{array}{c}
\hspace{-3cm}t_{0}~=~~~~~~~t_{s}(r)\\
\hspace{-2cm}lim~r\rightarrow 0\\\\
\hspace{1.1cm}=t_{i}+\frac{2}{3\sqrt{C_{0}}}~_{2}F_{1}[\frac{1}{2},b,b+1,z],~~~~~\left(z=\frac{g_{0}}{C_{0}(3-n)}\right)
\end{array}
\end{equation}

where in evaluating the limit we have used the series form of
$g(r)$  and $C(r)$ (from eq. (56)). Now if we restrict $n<3$ then
we have a comparative expression between $t_{ah}(r)$ and $t_{0}$
as

\begin{eqnarray*}
t_{ah}(r)-t_{0}=\left[-C_{0}^{-3/2}C_{1}~_{2}F_{1}[\frac{1}{2},b,b+1,z]+
\frac{(C_{0}g_{1}-C_{1}g_{0})}{(3-n)(9-2n)}C_{0}^{-5/2}~_{2}F_{1}[\frac{3}{2},b+1,b+2,z]\right]r
\end{eqnarray*}

\begin{equation}
+O(r^{2})-\frac{C_{0}^{3-n}g_{0}}{(3-n)(9-2n)}~_{2}F_{1}[\frac{1}{2},b,b+1,C_{0}^{3-n}~z]~r^{9-2n}+....~~....~,~~~~(n<3)
\end{equation}

Note that here $t_{0}$ is the time of formation of singularity at
$r=0$ while $t_{ah}(r)$ is the epoch at which a trapped surface is
formed at a distance $r$. Thus if trapped surface is formed at a
later instant than $t_{0}$ then it is possible that any light
signal from the singularity can reach an observer. As the
geometry of the present model is a class of spherical space-time
having different centres, so the condition for formation of NS
(or BH) will be same as TBL model. Therefore, $t_{ah}(r)>t_{0}$ is
the necessary condition for formation of naked singularity, while
to form black hole the sufficient condition is $t_{ah}(r)\le
t_{0}$. It should be mentioned that this criterion
for naked singularity is purely local.\\

Due to complicated form of equation (64) it is very difficult to
make a comparative study between $t_{ah}$ and $t_{0}$. Hence for
simplicity we choose $n=3/2$. Then the difference between $t_{ah}$
and $t_{0}$ has the form

\begin{equation}
t_{ah}(r)-t_{0}=\frac{2\left(C_{0}g_{1}-C_{1}g_{0}-g_{1}\sqrt{C_{0}}\sqrt{C_{0}-
\frac{2}{3}g_{0}}\right)}{3g_{0}\sqrt{C_{0}}\sqrt{C_{0}-\frac{2}{3}g_{0}}\left(\sqrt{C_{0}}+
\sqrt{C_{0}-\frac{2}{3}g_{0}}\right)}~r+O(r^{2})
\end{equation}

Hence in the present problem it is possible to have local naked
singularity or a black hole form under the conditions shown in the
following table (see Table I):\\

\newpage
\[
\text {TABLE-I}
\]
\[
\begin{tabular}{|l|l|r|r|r|}
\hline\hline \multicolumn{1}{|c|}{~~~Choice of the parameters~~~}
& \multicolumn{1}{c|}{~~~~Naked Singularity~~~~} &
\multicolumn{1}{c|}{~~~~Black hole~~~~}  \\
\hline\hline
&  &     \\
(i)~~ $g_{1}>0, C_{1}<0$  &  Always possible  &  Not possible \\

\hline
&  &     \\
(ii) ~$g_{1}<0, C_{1}>0$  &  Not possible &  Always possible \\

\hline
&  &     \\
(iii) $g_{1}>0, C_{1}>0$  &
$\frac{g_{1}}{C_{1}}>\frac{3}{2}\left(1+\sqrt{1-\frac{2}{3}k_{0}}\right)$
&
$\frac{g_{1}}{C_{1}}<\frac{3}{2}\left(1+\sqrt{1-\frac{2}{3}k_{0}}\right)$
\\

\hline
&  &    \\
(iV) $g_{1}<0, C_{1}<0$  &
$|\frac{g_{1}}{C_{1}}|<\frac{3}{2}\left(1+\sqrt{1-\frac{2}{3}k_{0}}\right)$
&
$|\frac{g_{1}}{C_{1}}|>\frac{3}{2}\left(1+\sqrt{1-\frac{2}{3}k_{0}}\right)$
\\

\hline\hline
\end{tabular}%
\]%
\newline

Here we note that for initial density gradient to be negative at
the centre (i.e., $\rho_{1}<0$) we must have
$(C_{1}-\frac{2g_{1}}{3})<0$ (for $\nu_{_{-1}}>-1$). In the first
case (i.e., $g_{1}>0, C_{1}<0$) we have negative definite
$\rho_{1}$ and there is always naked singularity as in the dust
model. Similarly in the second case (i.e., $g_{1}<0, C_{1}>0$),
$\rho_{1}$ is positive definite and we always get black hole same
as dust model. For the third and fourth cases (when $g_{1}$ and
$C_{1}$ are of same sign) both naked singularity (NS) and black
hole (BH) are possible for the restrictions given in the table I.
When both $g_{1}$ and $C_{1}$ are positive (third case) or
negative (fourth case) then for formation of NS $\rho_{1}$ is
negative but for BH case as there is no lower limit (or upper
limit) of $\frac{g_{1}}{C_{1}}$ (or $|\frac{g_{1}}{C_{1}}|$) so
$\rho_{1}>0$ or $\rho_{1}<0$ are possible. Further for
$g_{1}=C_{1}=0$ we have $\rho_{1}=0$ then we have similar
behaviour for the parameters ($g_{2}, C_{2}$). A diagrammatic
representation of $t_{ah}-t_{0}$ for variation of
$k_{0}(=g_{0}/C_{0})$ and $k_{1}(=g_{1}/C_{1})$ has been shown in
figures 1 and 2 for positive and negative $C_{1}$ respectively. In
both the figures the vertical positive region corresponds to NS
while the negative region stands for BH solution. Finally, we see
that if the initial density or pressure has opposite behaviour
(i.e., one increases and other decreases and vise-versa) near the
centre $r=0$ then we have similar character of the singularity as
in dust model i.e., pressure has no significant effect on the
singularity formation. On the other hand, if initial density and
pressure increase or decrease simultaneously near the centre then
even for negative density gradient at the centre it is possible to
have a BH formation at $r=0$, which is a distinct result in
compare to dust model. Therefore, we may conclude that
pressure tries to resist formation of NS.\\

\begin{figure}
\includegraphics[height=1.7in]{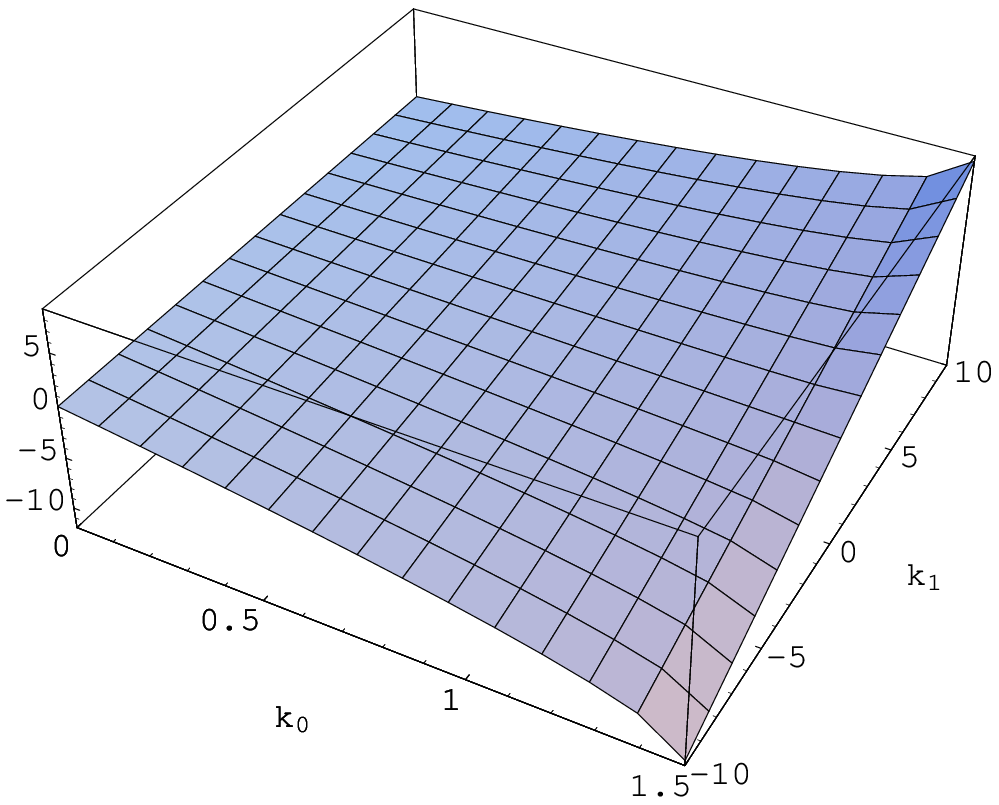}~~~
\includegraphics[height=1.7in]{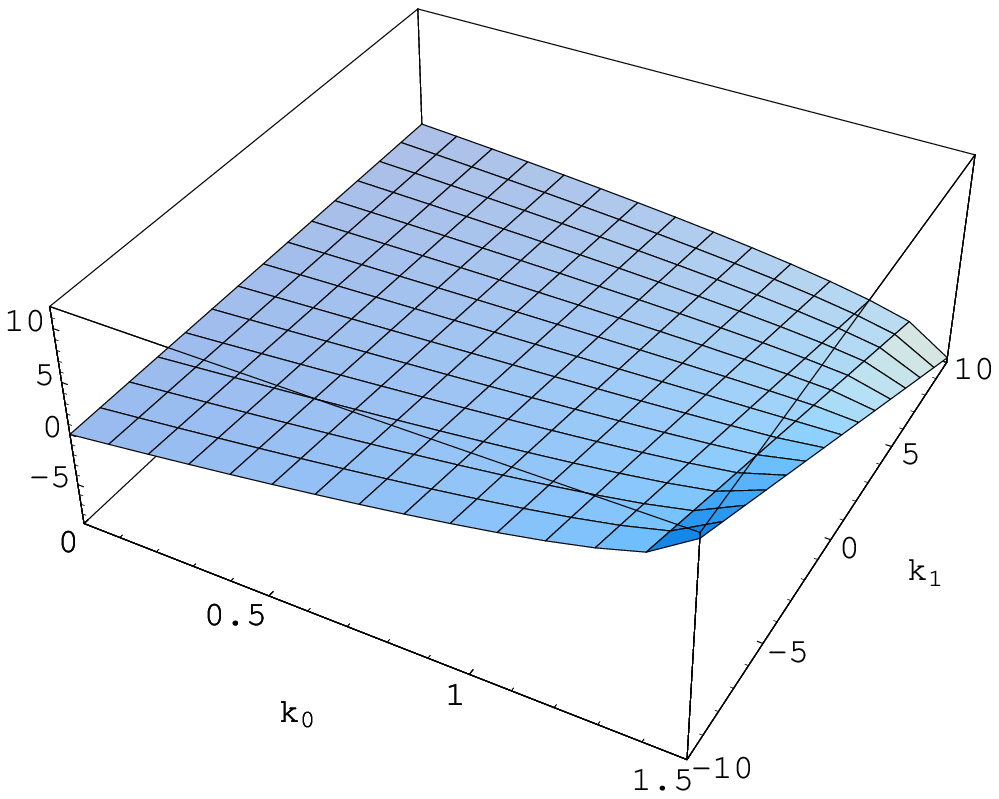}\\
\vspace{1mm}
Fig.1~~~~~~~~~~~~~~~~~~~~~~~~~~~~~~~~~~~~~~~~~~~Fig.2\\
\vspace{5mm} Figs. 1 and 2 show variation of $t_{ah}-t_{0}$ of
eq.(65) for the variation of $k_{0}(=g_{0}/C_{0})$ and
$k_{1}(=g_{1}/C_{1})$. Fig.1 corresponds to $C_{1}>0$ while Fig.2
corresponds to $C_{1}<0$. \hspace{1cm} \vspace{6mm}

\end{figure}

\subsection{\normalsize\bf{Study of Geodesics}}

For simplicity of calculation here we shall consider as before the
marginally bound case ($f(r)=0$) with $n=3/2$. Then $R(t,r)$ has
the explicit solution (choosing the initial time $t_{i}=0$) which
can be written in convenient form as

\begin{equation}
t(r)=\frac{2}{g(r)}\left[\sqrt{C(r)-\frac{2}{3}g(r)R^{3/2}}-\sqrt{C(r)-\frac{2}{3}g(r)r^{3/2}}\right]
\end{equation}

To examine whether the singularity at $t=t_{0}, r=0$ is naked or
not, we investigate whether there exist one or more outgoing null
geodesics which terminate in the past at the central singularity.
In particular, we shall concentrate to radial null geodesics
only.\\

First we assume that it is possible to have one or more such
geodesics and we choose the equation of the outgoing radial null
geodesic (ORNG) which passes through the central singularity in
the past as (near $r=0$)
\begin{equation}
t_{ORNG}=t_{0}+a~r^{\xi}
\end{equation}

to leading order in the ($t,r$) plane with $a>0, \xi>0$.\\

Now the expression for the singularity time (characterized by
$R(t_{s}(r),r)=0$) from (66) is
\begin{equation}
t_{s}(r)=\frac{2}{g(r)}\left[\sqrt{C(r)}-\sqrt{C(r)-\frac{2}{3}g(r)r^{3/2}}\right]
\end{equation}

and hence the time for central singularity is

\begin{equation}
t_{0}=\frac{2}{g_{0}}\left(\sqrt{C_{0}}-\sqrt{C_{0}-\frac{2}{3}g_{0}}\right)
\end{equation}

Here we choose for $C(r)$ and $g(r)$ as

\begin{equation}\begin{array}{c}
C(r)=C_{0}r^{3}+C_{k}r^{k+3}
\\\\g(r)=g_{0}~r^{3/2}+g_{_{l}}~r^{l+3/2}
\end{array}
\end{equation}

where $C_{0}, g_{0}$ are constants and $C_{k} (<0)$ and $g_{_{l}}
(<0)$ are the first non-vanishing term beyond $C_{0}$ and $g_{0}$
respectively. As a consequence the expression for $t_{s}(r)$
becomes

\begin{equation}
t_{s}(r)=t_{0}+\frac{C_{k}}{g_{0}}\left(\frac{1}{\sqrt{C_{0}}}-\frac{1}{\sqrt{C_{0}-
\frac{2}{3}g_{0}}}\right)r^{k}+\frac{2g_{_{l
}}}{g_{0}}\left(\frac{1}{3\sqrt{C_{0}-\frac{2}{3}g_{0}}}-
\frac{\sqrt{C_{0}}-\sqrt{C_{0}-\frac{2}{3}g_{0}}}{g_{0}}
\right)r^{l}+....
\end{equation}

we shall now study the following possibilities:\\
$$(i)~~k<l~,~~~~~~(ii)~~k>l$$

{\bf Case I} :~~  $k<l$ \\

Here for $t_{s}(r)$ we write
\begin{equation}
t_{s}(r)=t_{0}-\frac{C_{k}}{g_{0}}\left(\frac{1}{\sqrt{C_{0}-\frac{2}{3}g_{0}}}-\frac{1}{\sqrt{C_{0}}}
\right)r^{k},~~~~(C_{k}<0)
\end{equation}

Now comparing with the geodesic equation (67) we get the relations

\begin{equation}
(a)~~\xi>
k~~\text{or}~~~(b)~~\xi=k~~~\text{and}~~~a<-\frac{C_{k}}{g_{0}}\left(\frac{1}{\sqrt{C_{0}-\frac{2}{3}g_{0}}}-\frac{1}{\sqrt{C_{0}}}
\right)
\end{equation}

When $\xi>k$ then near $r=0$ the solution for $R$ simplifies to

\begin{equation}
R=r\left[1-\frac{3}{8}g_{0}t^{2}-\frac{3}{2}~t\left(\sqrt{C_{0}-\frac{2}{3}g_{0}}+\frac{C_{k}r^{k}}{2\sqrt{C_{0}-
\frac{2}{3}g_{0}}}\right) \right]^{2/3}
\end{equation}

Further for the given metric an ORNG should satisfy

\begin{equation}
\frac{dt}{dr}=R'+R\nu'
\end{equation}

To examine the feasibility of the null geodesic starting from the
singularity, we combine equations (67) and (74) in equation (75)
and we get (upto leading order in $r$)

\begin{equation}
a\xi
r^{\xi-1}=\left(1+\nu_{_{-1}}+\frac{2k}{3}\right)\left[-\frac{3C_{k}t_{0}}{4\sqrt{C_{0}-
\frac{2}{3}g_{0}}}\right]^{2/3}r^{\frac{2k}{3}}
~,~~~~~(\nu_{_{-1}}\ne 0)
\end{equation}

which implies
\begin{equation}
\xi=1+\frac{2k}{3}
~~~\text{and}~~~a=\frac{1}{\xi}\left(1+\nu_{_{-1}}+\frac{2k}{3}\right)\left[-\frac{3C_{k}t_{0}}
{4\sqrt{C_{0}-\frac{2}{3}g_{0}}}\right]^{2/3}
\end{equation}

As $\xi>k$, so from (77) $k<3$. Since $k$ is an integer, we could
have

\begin{equation}
\begin{array}{c}
k=1,~\xi=\frac{5}{3}\\\\
\text{or}\\\\\
k=2,~\xi=\frac{7}{3}
\end{array}
\end{equation}

On the other hand for $\xi=k$, as before we get $k=3$ and

\begin{eqnarray*}
a=\frac{1}{3}\left[-\frac{3}{4}\left(ag_{0}t_{0}+2a\sqrt{C_{0}-\frac{2}{3}g_{0}}+
\frac{C_{k}t_{0}}{\sqrt{C_{0}-\frac{2}{3}g_{0}}}\right)\right]^{-1/3}~\times
\end{eqnarray*}

\begin{equation}
\left[-\frac{3}{4}
\left\{(1+\nu_{_{-1}})\left(ag_{0}t_{0}+2a\sqrt{C_{0}-\frac{2}{3}g_{0}}\right)+
\frac{\left(2+\nu_{_{-1}}+\frac{4k}{3}\right)C_{k}t_{0}}{\sqrt{C_{0}-\frac{2}{3}g_{0}}}
\right\} \right]
\end{equation}
\\\\\\\\

{\bf Case II}: $k>l$\\

In this case
\begin{equation}
t_{s}(r)=t_{0}-\frac{2g_{_{l}}}{g_{0}^{2}}\left(\sqrt{C_{0}}+\frac{g_{0}-3C_{0}}{3\sqrt{C_{0}-\frac{2}{3}g_{0}}}
\right)r^{l}
\end{equation}

Now matching the geodesic equation as above we get

\begin{equation}
(a)~~\xi>
l~~\text{or}~~(b)~~\xi=l~~\text{and}~~a<-\frac{2g_{_{l}}}{g_{0}^{2}}\left(\sqrt{C_{0}}+\frac{g_{0}-3C_{0}}{3\sqrt{C_{0}-\frac{2}{3}g_{0}}}
\right)
\end{equation}

Then for $\xi>l$ we get

\begin{equation}\begin{array}{c}
l=1,~ \xi=\frac{5}{3} ~~~\text{or}~~~l=2,~ \xi=\frac{7}{3}~;\\\\
a=\frac{1}{\xi}\left(1+\nu_{_{-1}}+\frac{2l}{3}\right)\left[-\frac{3g_{_{l}}t_{0}}{8}\left(t_{0}-
\frac{4}{3\sqrt{C_{0}-\frac{2}{3}g_{0}}}\right) \right]^{2/3}
\end{array}
\end{equation}

and for $\xi=l$

\begin{eqnarray*}
l=3,~~~~a=\frac{1}{3}\left[-\frac{3}{8}\left(g_{0}t_{0}^{2}+2at_{0}g_{0}\right)-\frac{3}{2}~a\sqrt{C_{0}-
\frac{2}{3}g_{0}}+\frac{g_{_{l}}t_{0}}{2\sqrt{C_{0}-\frac{2}{3}g_{0}}}
\right]^{-1/3}~\times
\end{eqnarray*}

\begin{equation}
\left[-\frac{3}{8}\left\{\left(1+\nu_{_{-1}}+\frac{2l}{3}\right)g_{0}t_{0}^{2}+2at_{0}
(1+\nu_{_{-1}})g_{0}\right\}-\frac{3}{2}~a(1+\nu_{_{-1}})\sqrt{C_{0}-
\frac{2}{3}g_{0}}+\frac{\left(1+\nu_{_{-1}}+\frac{2l}{3}\right)g_{_{l}}t_{0}}{2\sqrt{C_{0}-\frac{2}{3}g_{0}}}
\right]
\end{equation}
\\

We note that the expressions for `$a$' is very complicated both
in equations (79) and (83). So no definite conclusion is possible
on the role of pressure in determining in the final state of collapse by ORNG.\\

\section{\normalsize\bf{Discussions and Concluding Remarks}}

An extensive analysis of the four dimensional Szekeres model has
been done for the matter containing pressure. When matter is in
the form of perfect fluid then the isotropic pressure turns out to
be a function of time only while the matter density is a function
of all the four space-time variables. In this case, assuming a
polynomial form for pressure, cosmological solutions have been
obtained and their asymptotic behaviour have been studied. Both in
quasi-spherical and quasi-cylindrical model the solution
approaches isotropy along fluid world line as
$t\rightarrow\infty$.\\

Secondly, for the matter with tangential stress only, solutions
are possible for quasi-cylindrical model. Here both the tangential
stress and the matter density turns out to be a function of $t$
and $r$ only. The scale factor $R$ has parametric solution as for
dust model and does not depend on the tangential stress. However,
choosing the parameter $C_{1}=0$, $R$ has a power law solution and
it is possible to have a complete solution if we assume the
tangential stress proportional to $t^{-2}$.\\

Lastly, gravitational collapse has been studied in details for
anisotropic pressure (i.e., both radial and tangential pressures
are non-zero and distinct) in quasi-spherical model. Here we have
to assume the radial pressure as a function of $r$ and $t$ of the
form (see eq. (50)) $p_{r}=g(r)/R^{n}$. Also to solve the
differential equation in $R$ (see eq. (53)) we consider only the
marginally bound case (i.e., $f=0$) only. Then equation (64) shows
a comparative study between the time of formation of trapped
surface and the time of formation of central singularity. To
simplified further, we choose $n=3/2$ and detailed analysis has
been done using equation (65). Table I shows all possibilities for
the parameters involved in the expression. If the initial density
gradient at the centre is positive definite (or negative definite)
then as in dust case we have definitely a black hole (or naked
singularity) as the final state of collapse. But when $\rho_{1}$
has no definite sign (as in third and fourth cases) then for black
hole solution it is possible to have negative density gradient at
the centre initially. In fact, near the singularity if the initial
density and pressure has identical behaviour (i.e., increase or
decrease simultaneously) then even with negative density gradient
(initially at the centre) we can have black hole as the end state
but if the initial density and pressure has opposite behaviour
(i.e., one decrease while other increases and vice versa) then we
have identical character as in dust case. So we conclude that
pressure tries to resist the formation of naked singularity.
Finally we have studied the geodesics to examine whether it is
possible to have any future directed non space-like geodesic
terminating in the past at the singularity. For simplicity, we
have considered only radial null geodesic and it is found that the
end state of collapse is characterized by the coefficients of the
series expansion of initial density and pressure (radial). Due to
complicated expressions we can not definitely characterize the
role of pressure. Therefore, in the context of local visibility,
we say that pressure tries to cover the singularity.\\\\

{\bf Acknowledgement:}\\

The authors are thankful to IUCAA for worm hospitality where the
major part of the work has been done. One of the authors (U.D) is
thankful to CSIR (Govt. of India)
for awarding a Senior Research Fellowship.\\

{\bf References:}\\
\\
$[1]$  O. Heckmann, E. Sch$\ddot{\text{u}}$cking; in : L. Witten
(Ed.): Gravitation, an Introduction
to current research, NewYork, Wiley (1962).\\
$[2]$  P. Szekeres, {\it Commun. Math. Phys.} {\bf 41} 55 (1975).\\
$[3]$ P. S. Joshi and I. H. Dwivedi, \textit{Commun. Math. Phys.}
\textbf{166 } 117 (1994).\\
$[4]$ P. S. Joshi and I. H. Dwivedi,\textit{Class. Quantum Grav.}
\textbf{16 } 41 (1999).\\
$[5]$ K. Lake, \textit{Phys. Rev.Lett.}\textbf{68} 3129 (1992).\\
$[6]$ A. Ori and T. Piran, \textit{Phys. Rev. Lett.} \textbf{59} 2137 (1987).\\
$[7]$ T.Harada, \textit{Phys. Rev. D} \textbf{58} 104015
(1998).\\
$[8]$ P.S. Joshi, \textit{Global Aspects in Gravitation and
Cosmology, }(Oxford Univ. Press, Oxford, 1993).\\
$[9]$ H. Muller zum Hagen, P. Yodzis and H. Seifert, {\it Commun.
Math. Phys.} {\bf 37} 29 (1974).\\
$[10]$ L. Herrera and N. O. santos, {\it Phys. Rep.} {\bf 286} 53
(1997).\\
$[11]$ M. Celerier and P. Szekeres, {\it gr-qc}/0203094; R.
Giambo', F. Giannoni, G. Magli and P. Piccione, {\it
gr-qc}/0204030.\\
$[12]$ T. Harada, H. Iguchi and K. Nakao, {\it Prog. Theor. Phys.}
{\bf 107} 449 (2002).\\
$[13]$ R. Goswami and P. S. Joshi, {\it Class. Quantum Grav.} {\bf
19} 5229 (2002).\\
$[14]$ K. S. Thorne, in \textit{Magic Without Magic}: \textit{John
Archibald Wheeler}, Ed. Klauder J (San Francisco: W. H. Freeman
and Co. 1972).\\
$[15]$ S. L. Shapiro and S. A. Teukolsky, \textit{Phys. Rev.
Lett.} \textbf{ 66} 994 (1991).\\
$[16]$ T. Nakamura, M. Shibata and K.I.Nakao, \textit{Prog. Theor.
Phys.}
\textbf{89} 821 (1993) .\\
$[17]$ C. Barrabes, W. Israel and P. S. Letelier, \textit{Phys.
Lett. A}\textbf{160} 41 (1991); M. A. Pelath, K. P. Tod and R. M.
Wald, \textit{Class. Quantum Grav.} \textbf{15} 3917 (1998).\\
$[18]$ T. Harada, H. Iguchi and K.I. Nakao, \textit{Phys. Rev. D}
\textbf{58} 041502 (1998) .\\
$[19]$ H. Iguchi, T. Harada and K.I. Nakao, \textit{Prog. Theor.
Phys.}\textbf{101} 1235 (1999); \textit{Prog. Theor. Phys.}
\textbf{103} 53 (2000).\\
$[20]$ P. Szekeres, \textit{Phys. Rev. D} \textbf{12} 2941
(1975).\\
$[21]$ U. Debnath, S. Chakraborty and J. D. Barrow, {\it Gen. Rel.
Grav.} {\bf 36} 231 (2004); U. Debnath and S. Chakraborty, {\it
JCAP}~ {\bf 05} 001 (2004).\\
$[22]$  P. Szekeres, {\it Commun. Math. Phys.} {\bf 41} 55 (1975).\\
$[23]$  S. Chakraborty and U. Debnath, {\it Int. J. Mod. Phys. D} {\bf 13} 1085 (2004)).\\
$[24]$ Note that equation (22) has many possible solutions of
which one is given in equation (24). Another solution can be taken
as ~$e^{-\nu}=A_{1}(r) \text{sin}(\lambda x)+A_{2}(r)
\text{cos}(\lambda x)+B_{1}(r) \text{sinh}(\lambda y)+B_{2}(r)
\text{cosh}(\lambda y)$ with
$\lambda^{2}(A_{1}^{2}+A_{2}^{2}+B_{1}^{2}-B_{2}^{2})=f(r)-1$.\\
$[25]$  D. A. Szafron, {\it J. Math. Phys.}
{\bf 18} 1673 (1977).\\
$[26]$  D. A. Szafron and J. Wainwright {\it J. Math. Phys.}
{\bf 18} 1668 (1977).\\

\end{document}